# Lorentz transformations: Einstein's derivation simplified[1]

## Bernhard Rothenstein[1] and Stefan Popescu[2]

1) Politehnica University of Timisoara, Physics Department,
Timisoara, Romania brothenstein@gmail.com
2) Siemens AG, Erlangen, Germany stefan.popescu@siemens.com

**Abstract**. *We show that the Lorentz transformations for the space-time coordinates of the same event are a direct consequence of the principle of relativity and of Einstein's distant clocks synchronization procedure. In our approach, imposing the linear character of the Lorentz transformations we guess that the transformation equation for the space coordinate has the form $x=ax'+cbt'$. Imposing the condition that it accounts for the time dilation relativistic effect and taking into account the fact that due to the clock synchronization à la Einstein the space-time coordinates of the same event in the two frames are related by $x=ct$ and $x'=ct'$, we find out expressions for a and b. Dividing the transformation equation for the space coordinate by c we obtain the transformation equation for the time coordinate $t=at'+bc^{-1}x'$. Combining the two transformation equations we obtain directly the inverse Lorentz transformations.*

Those who have studied Einstein's special relativity theory know that everything there is the result of his two postulates and of the distant clock synchronization procedure that he proposed.

The concept of **event** is fundamental in physics being defined as any physical occurrence that could take place at a given point in space and at a given time. The notation $E(x, y=0, t)$ defines an event that takes place at a given point $M(x,0)$ of the OX axis of the K(XOY) inertial reference frame when the clock $C(x, y=0)$ located at that point reads t. In order to be operational the different clocks of that frame, located along the OX axis should display the same running time. Einstein satisfied that condition proposing the synchronization procedure shown in Figure 1.

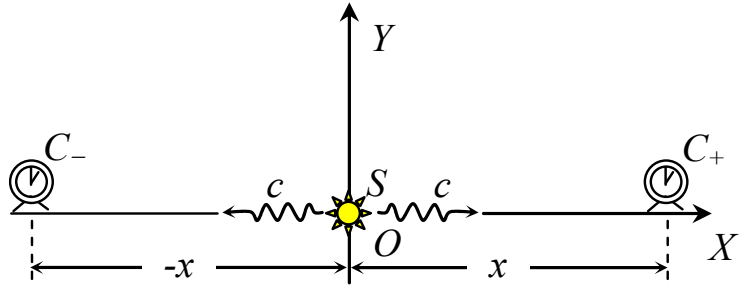

***Figure 1***. *The synchronization of clocks $C_+(x,0)$ and $C_-(-x,0)$ following a clock synchronization procedure as proposed by Einstein.*

Clock $C_0(0,0)$ located at the origin O is ticking and when it reads a zero time the source of light $S(0,0)$ located in front of it emits short light signals in the positive and in the negative directions of the OX axis. Clocks $C_+(0,0)$ and $C_-(0,0)$ are initially stopped and fixed to display a time $t=x/c$. The light signals arriving at the corresponding clocks start them and from that very moment the clocks display the same running time. The events associated with the synchronization of clocks $C_0$, $C_+$ and $C_-$ are $E_0(0,0,t)$, $E_+(x,0,t)$ and $E_-(-x,0,t)$ respectively. It is obvious that their space-time coordinates are related by

$$x = \pm ct \qquad (t > 0) \qquad (1)$$

or by

$$x^2 - c^2 t^2 = 0 \qquad (2)$$

Special relativity becomes involved when we consider a second inertial reference frame K'(X'O'Y') in the standard arrangement with the K(XOY) reference frame, K' moving with constant velocity V in the positive direction of the overlapped OX(O'X') axes. The events associated with the synchronization of the clocks in K' are $E'_0(0,0,t)$, $E'_+(x',0,t')$ and $E'_-(-x',0,t')$.

The clocks $C'_0(0,0)$, $C'_+(x',0)$ and $C'_-(0,0)$ of that frame are synchronized following the same procedure as in K and we have obviously

$$x'^2 - c'^2 t'^2 = 0. \qquad (3)$$

Equating (2) and (3) we obtain

$$x^2 - c^2 t^2 = x'^2 - c^2 t'^2. \qquad (4)$$

Because at the origin of time the origins of K and K' are located at the same point in space we can consider that $\Delta x=x-0$, $\Delta t=t-0$, $\Delta x'=x'-0$ and $\Delta t'=t'-0$ presenting (4) as

$$(\Delta x)^2 - c^2(\Delta t)^2 = (\Delta x')^2 - c^2(\Delta t')^2 \qquad (5)$$

Equation (5) is a starting point in Einstein's derivation of the Lorentz transformations[1] which establish a relationship between the space-time coordinates of events E(x,0.t) and E'(x',0,t').

Relativists consider that one event E(x,0,t) detected from the K frame and an event E'(x',0,t') detected from the K' frame represent the **same** event if they take place at the same point in space when the clocks C(x,0) and C'(x',t') located at that point read t and t' respectively. The Lorentz transformations establish a relationship between the coordinates of events E(x,0,t) and $E'(x',0,t)$ defined above and considered to represent the same event. We derive them in two steps. Figure 2 presents the relative position of the reference frames K and K' as detected from the K frame when its clocks read t.

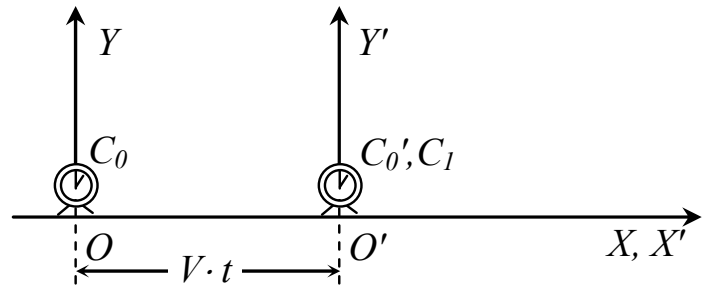

*Figure 2*. *Scenario that leads to the time dilation formula. We compare the readings of clocks $C_1(x=Vt,0)$ and $C'_0(0,0)$ when they are located at the same point in space as detected from the inertial reference frame K.*

When the clock $C'_0(0,0)$ is reading t' it is located in front of a clock $C_1$(x=Vt,0) reading t. The problem is to establish a relationship between Δt and Δt'. The clock $C'_0(0,0)$ being in a state of rest in K' we have in its case Δx'=0. The position of clock $C'_0(0,0)$ is defined in K by Δx=VΔt, the change in the reading of clock $C_1$(x=Vt,0) being $\Delta t$. The events involved are E(x=Vt,t) in K and (x'=0,t') in K'. Imposing the condition (4) that relates correctly their space-time coordinates we obtain

$$t = \frac{t'}{\sqrt{1-\frac{V^2}{c^2}}} \qquad (6)$$

which relates the readings of the two clocks when they are located at the same point in space, equation

$$\Delta t = \frac{\Delta t'}{\sqrt{1-\frac{V^2}{c^2}}} \qquad (7)$$

relating the changes in theirs readings.[2] It is of essential importance to make a net distinction between the ways in which the time intervals Δt and Δt' are measured.

The time interval Δt is measured as a difference between the reading t of clock C(x=Vt,0) and the reading t=0 for clock $C_0$(0,0) when the moving clock $C'_0(0,0)$ passes in front of them respectively. Relativists call a time interval measured under such conditions **coordinate time interval.** The time interval Δt' is measured as a difference between the readings of the same clock $C'_0(0,0)$ when it passes in front of clock $C_1$(x=Vt,0), (t') and when it passes in front of clock $C_0$(0,0) (t'=0). A time interval measured under such conditions is called **proper time interval**. As we see (7) relates a coordinate time interval measured in the reference frame K and a proper time interval measured in K. Because Δt>Δt' relativists say that a **time dilation effect** takes place. If we consider the same experiment from the inertial reference frame K' then we see that observers of that reference frame measure a coordinate time interval whereas observers from K measure a proper time interval related by (7). Figure 3 presents the relative positions of the reference frames K and K' when all the clocks of the first frame read t.

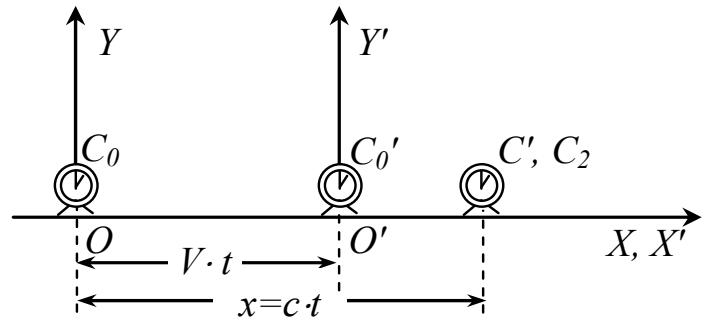

***Figure 3***. *Scenario for deriving the Lorentz transformations*

We concentrate on the clocks $C_2$(x=ct,0) of the K frame which reads t and the clock C'(x'=ct',0) of the K' frame which reads t' when they are located at the same point in space. The relations x=ct and x'=ct' are a direct consequence of Einstein's clock synchronization procedure. The problem is to find out a relationship between the space-time coordinates of events $E_2$(x=ct,0,t) and $E'_2(x'=ct',0,t')$ associated with the space coincidence of the two clocks. Such a relationship is known as **Lorentz transformation for** **the space-time coordinates of the same event. It should be linear because if the motion is uniform in one of the involved reference frames it should be uniform in all other inertial reference frames moving uniformly relative to it. Besides that a transformation equation applied to (5) should confirm the invariance of the physical quantities it equates.**

Encouraged by Galileo's transformation equations[3]

$$x = x' + Vt' \tag{8}$$
$$x' = x - Vt \tag{9}$$
$$t = t' \tag{10}$$

we guess that in Einstein's special relativity theory, one of the transformation equations should have the shape

$$x = ax' + cbt' \tag{11}$$

where *a* and *b* represent factors which, due to the linear character of a transformation equation, could depend on the relative velocity V but not on the space-time coordinates of the involved events. In order to find them we impose the condition that it should correctly relate the space-time coordinates of events E(x=Vt,0,t) and E'(x'=0,0,t') and of events E'(x'=-Vt',0,t'), E(0,0,t) we have defined deriving the formula which accounts for the time dilation effect. In the case of the first pair of events (11) works as

$$Vt = bct' = bct\sqrt{1 - \frac{V^2}{c^2}} \tag{12}$$

wherefrom we obtain

$$b = \frac{\frac{V}{c}}{\sqrt{1 - \frac{V^2}{c^2}}} = \beta c^{-1}\gamma(V). \tag{13}$$

In the case of the second pair of events (11) works as

$$0 = -aVt' + cbt' \tag{14}$$

resulting that

$$a = \gamma(V) \tag{15}$$

(11) becoming

$$x = \gamma(V)(x' + Vt'). \tag{16}$$

Dividing both sides of (16) by *c* and taking into account that all the involved clocks are synchronized à la Einstein (t=x/c,t'=x'/c) we obtain

$$t = \gamma(V)(t' + \beta c^{-1}x') \tag{17}$$

Combing (16) and (17) we obtain with some algebra

$$x' = \gamma(V)(x - Vt) \tag{18}$$

$$t' = \gamma(V)(t - \beta c^{-1}x). \tag{19}$$

Equations (16) and (17) are known as the **inverse Lorentz transformations** whereas equations (18) and (19) are known as the **direct Lorentz transformations.**

Compared with Einstein's derivation and with other derivations we found in the literature of this subject, our derivation presents the advantage that it is shorter, revealing the fact that the Lorentz transformations are a direct consequence of the two relativistic postulates and of the clock synchronization procedure proposed by Einstein.

The Lorentz transformations become more transparent if we present them as a function of changes in the space-time coordinates of the same event. Equations (16) and (17) become

$$\Delta x = \gamma(V)(\Delta x' + V \Delta t') \tag{20}$$

and

$$\Delta t = \gamma(V)(\Delta t' + V c^{-2} \Delta x'). \tag{21}$$

The way in which the transformation equations were derived ensures the fact that they account for the time dilation effect. They account in a transparent way for the addition law of relativistic velocities. Consider a particle that starts to move at t=t'=0 from the common origin of K and K' with speed $u_x$ relative to K and with speed $u'_x$ relative to K'. After a time of motion t the particle generates the event E(x=$u_x$t,0,t) as detected from K and E'($x'=u'_x t', 0, t'$) when detected from K'. In accordance with the Lorentz transformations we have

$$\Delta x = \Delta t' \frac{u'_x + V}{\sqrt{1 - \frac{V^2}{c^2}}} \tag{22}$$

and

$$\Delta t = \Delta t' \frac{1 + \frac{V u'_x}{c^2}}{\sqrt{1 - \frac{V^2}{c^2}}}. \tag{24}$$

By definition we have

$$u_x = \frac{\Delta x}{\Delta t} = \frac{u'_x + V}{1 + \frac{V u'_x}{c^2}} \tag{25}$$

an equation known as **the addition law of relativistic velocities.** It is worth to underline that Δx and Δx' represent proper lengths whereas Δt and Δt' represent coordinate time intervals.

**Conclusion**

Many authors derived the Lorentz transformation equations by following the way first suggested by Einstein.[4,5] All of them start with the guessed shape of the transformation equations for the space and time coordinates. In the one dimensional case this approach involves four unknown factors. Our approach involves only two unknown factors instead.